\documentclass[twocolumn,conference, lettersize]{IEEEtran}
\usepackage{graphicx, amsmath, amsfonts, epsfig, ntheorem,
  flushend, subfigure, color}
  \usepackage{nicefrac}
\usepackage{algorithm}
\usepackage{algorithmic}
  \usepackage{psfrag}
  \usepackage{manfnt}
    \usepackage{cite}

\usepackage{dsfont}

\newif\iftodo   
\todofalse 
\newif\iftodoshort  
\todoshortfalse
\newcommand{\todo}[1]
{\iftodo
  \iftodoshort
  \reversemarginpar\marginpar{\textdbend \sf\scriptsize ToDo}
  \else
  \\[.1em]\par
  \reversemarginpar\marginpar{\textdbend \sf\scriptsize ToDo}
  \fbox{\begin{minipage}[c]{0.95\linewidth}
       {\color{blue}{\small #1}}
    \end{minipage}
  }
  \\[.1em]\par
  \fi 
  \fi}

\theoremstyle{plain}
\theoremheaderfont{\normalfont\bfseries}\theorembodyfont{\normalfont}
\theoremsymbol{}
\theoremseparator{:}

\theoremheaderfont{\normalfont}\theorembodyfont{\normalfont}
\theoremstyle{nonumberplain}
\theoremseparator{}
\theoremsymbol{}

\theoremstyle{plain}
\theoremseparator{:}
\theoremheaderfont{\normalfont}\theorembodyfont{\normalfont}

\theoremstyle{plain}
\theoremheaderfont{\normalfont\bfseries}\theorembodyfont{\normalfont}
\theoremsymbol{}
\theoremseparator{:}

\theoremstyle{plain}
\theoremheaderfont{\normalfont\bfseries}\theorembodyfont{\normalfont}
\theoremsymbol{}
\theoremseparator{:}

\theoremstyle{plain}
\theoremheaderfont{\normalfont\bfseries}\theorembodyfont{\normalfont}
\theoremsymbol{}
\theoremseparator{:}

\begin{document}

\title{Bidirectional multi-pair network with a MIMO relay: Beamforming strategies and lack of duality}

\author{Aydin Sezgin$^1$, Holger Boche$^2$,  and Amir Salman Avestimehr$^3$}

\maketitle
\footnotetext[1]{Emmy-Noether Research Group on Wireless Networks,  Ulm University, -TAIT-, Albert-Einstein-Allee 43, 89081 Ulm, Germany, aydin.sezgin@uni-ulm.de. The work of A.Sezgin is supported by the
  Deutsche Forschungsgemeinschaft (DFG) under grant Se1697/3-1.}
  \footnotetext[2]{Technical University Berlin, Germany, Chair for Mobile Communications,  Einsteinufer 37, 10587 Berlin, Germany , boche@hhi.de}
  \footnotetext[3]{
  Cornell University, School of Electrical and Computer Engineering,  325 Frank H.T. Rhodes Hall,Ithaca, NY 14853, avestimehr@ece.cornell.edu}

\begin{abstract}
We address the problem of a multi-user relay network, where multiple single-antenna node pairs want to
exchange information by using a multiple antenna relay node. Due to the half-duplex constraint of the relay,
the exchange of information takes place in two steps. In the first step, the nodes transmit their data to the relay,
while in the second step, the relay is broadcasting the data by using linear and non-linear precoding strategies.
We focus on the second step in this paper. We first consider the problem of maximizing the overall rate achievable
using linear and dirty-paper type precoding strategies at the relay. Then, we consider minimizing the total power at
the relay subject to individual $\mathsf{SINR}$ constraints using the same strategies at the relay.
We show that the downlink-uplink duality does not hold for the setup considered here, which is a somewhat
surprising result. We also show that the beamforming strategy which is optimal in the single-pair case
performs very well in the multi-pair case for practically relevant $\mathsf{SNR}$. The results are
illustrated by numerical simulations.
\end{abstract}
%

\section{Introduction}
Cooperative communication plays a major role in future wireless ad-hoc-networks as well as in cellular systems such as LTE-Advanced.  Cooperation can take place between base stations or mobile stations directly as well as via a relay station. Relay stations can hereby vary in the capability and complexity. They can be equivalent to base stations, i.e. with connection to the backbone, or as simple as a mobile station itself and variations between those extremes.
The relays are then deployed within a cell or network in order to extend coverage or increase the battery life of mobile nodes.
There is huge activity in the research community analyzing different aspects of relay networks.
However, most of the work in multi-user relay networks is focused on the uni-directional case as e.g. in~\cite{KramerGastparGupta,MaricDaboraGoldsmith} and is often limited to the case where multiple source nodes transmit data to their receiving counterpart(s) by exploiting the relay.

Alternatively, the relays can also be used in order to exchange information between two nodes. The exchange of information between two nodes, often referred to as bidirectional communication, has been analyzed already by Shannon in~\cite{ShannonInt}. Some achievable rate regions for the bidirectional relay channel using different strategies at the relay, such as decode-and-forward, compress-and-forward, and amplify-and-forward, have been analyzed in~\cite{RankovISIT, GunduzAllerton08}. The performance of the bidirectional relay channel using superposition or network coding type strategies was analyzed in~\cite{RafaelOechtBiDiBroadISIT, OSBB08bcro,Bauchspawc2007}. In~\cite{AvestimehrSezgin}, some work has been done on near optimal relaying strategies as well as approximating the capacity region of the noisy (Gaussian) bidirectional relay channel. There, the relay uses a \emph{equation-forwarding} scheme, in which the relay re-orders the received superposed
signals, quantize and forwards them. It was shown that this scheme achieves rates which are within $3$ bits of the cut-set upper bound. The scaling of the capacity region with multiple antennas and multiple relays for two-way relaying is considered in~\cite{VazeHeath}.

For practical reasons, the relays involved in such cooperative communication systems often have an half-duplex constraint, i.e. they are not able to transmit and receive at the same time or at the same frequency simultaneously.
Due to this half-duplex constraint of the relay, the exchange of information takes place in two steps. In the first multiple-access step, the nodes transmit their data to the relay, while in the second step, the relay is broadcasting data to the nodes.
The capacity region of the broadcast step of this bidirectional relay channel was recently characterized in~\cite{OSBB08bcro}.

Naturally, in a network there are multiple transceiver pairs which would like to exchange information by using the same relay.
Obviously, this causes interference and thus in order to to have a reliable communication, interference management strategies are required~\cite{GunduzITMulticast,GunduzMultiWayIT}.
For example, in~\cite{ChenYener,ChenYenerCISS} interference management is performed by allocating spread sequences and power
efficiently to the communicating transceiver pairs.  In~\cite{AvestimehrITW09} the  capacity region of an
deterministic~\cite{AvestimehrThesis} multi-pair bidirectional relay network is characterized  and it was shown that
the cut-set upper bound is tight.  The results were then extended to the Gaussian case in~\cite{SezKhaAveHasISIT09} and an approximate capacity characterization is provided.

The research on multi-pair bidirectional relaying is so far focused on single-antenna nodes.
In this work, the relay is equipped with multiple antennas, while the nodes are equipped with a single antenna.
With the generalization to a MIMO relay we will face new and interesting challenges, as we will see later on.
Furthermore, the focus in this paper is on the broadcast phase of the bidirectional relay network.
In more details, we are investigating the performance of two transmit strategies namely
linear and nonlinear (dirty-paper~\cite{CostaDirty}) precoding performed by the relay in the broadcast phase.
Our contribution is thus the performance analysis of these strategies in terms of achievable rate region using different beamformers:
\begin{itemize}
\item single-pair beamforming by treating the interference of the other (or remaining) pairs as noise
\item beamformers obtained after solving a related power minimization problem with $\mathsf{SINR}$ constraints 
using a relaxation method
\item an exhaustive search over a huge set of randomly generated beamformers.
\end{itemize}
Note that the downlink in the second phase is fundamentally different then the broadcast or multicast channel, since the receiving nodes have side information in form of their own messages conveyed in the first phase.

The remainder of the paper is organized as follows. In section~\ref{sec:Model}, the system model is introduced, followed by section~\ref{sec:Rate} and~\ref{sec:Power}, in which the transmit strategies and the optimization problems are described. In addition, the lack of duality is discussed in this section. The results are then illustrated in section~\ref{sec:Illus}, followed by some concluding remarks in section~\ref{sec:Conc}.

\section{System model}\label{sec:Model}
\begin{figure}
\begin{center}
    \subfigure[First hop]{\label{fig:FirstHop}
    \includegraphics[width=4cm]{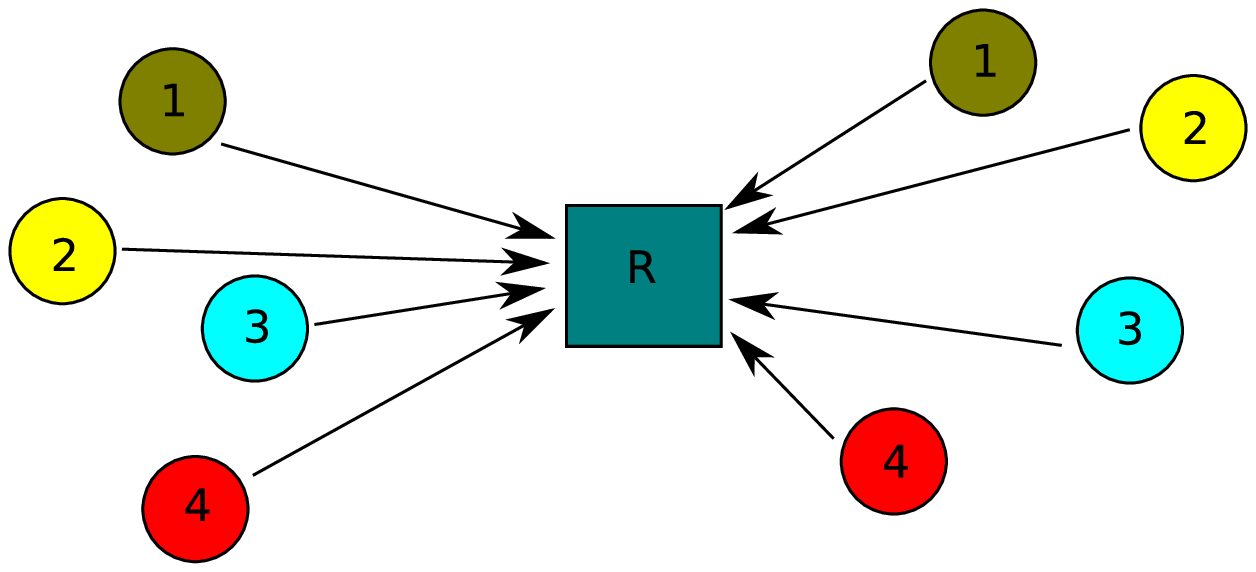}}
    \subfigure[Second hop.]{\label{fig:SecHop}
    \includegraphics[width=4cm]{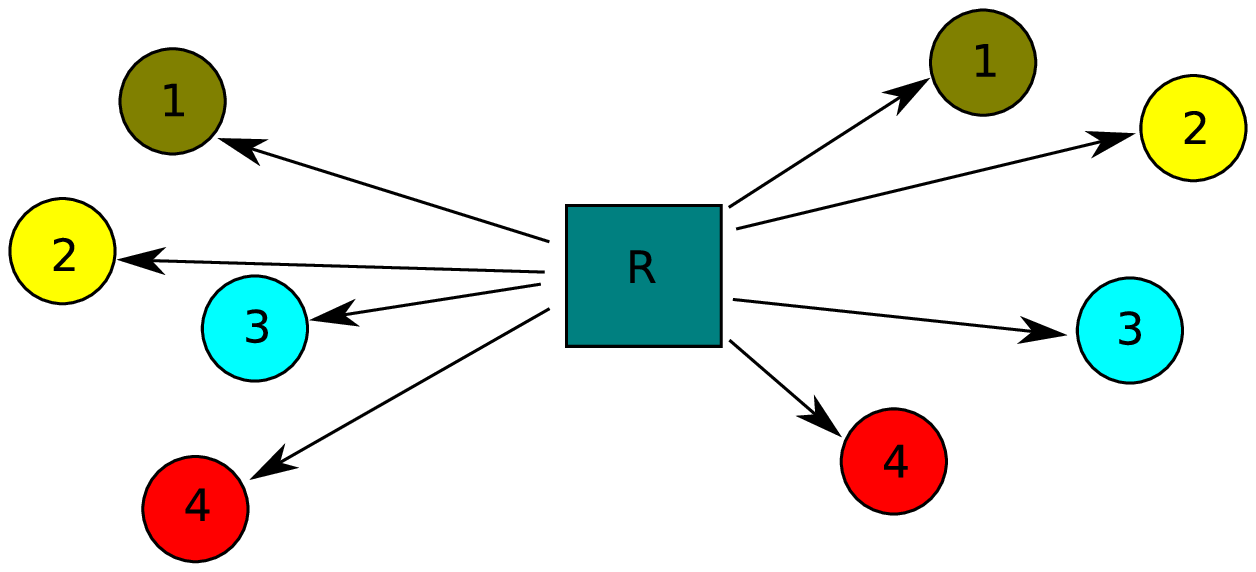}}
\end{center}
    \caption{Bidirectional relaying with multiple pairs. Pairs are indicated by equal numbers.\label{fig:ilid}}
\end{figure}

Suppose that there is a group of single-antenna transceiver pairs communicating to each other in a bidirectional way by exploiting a relay equipped with $N$ antennas as shown in Fig.~\ref{fig:ilid}. We assume that there are in total $N$ pairs of nodes. Each node is denoted by the tuple $(i,k)$, with $1\leq i\leq N$ and $k=\{1,2\}$, where the first index $i$ identifies the pair and the second index $k$ identifies which node of this node-pair is meant. Communication takes place in two hops. In the first hop, the multiple access step, the nodes transmit their data to the relay station. The relay station processes the received data and in the second hop, the broadcasting step, the data is forwarded to the nodes.

\begin{figure}
\begin{center}
\psfrag{W11}{$W_{(1,1)}$}
\psfrag{W12}{$W_{(1,2)}$}
\psfrag{W21}{$W_{(2,1)}$}
\psfrag{W22}{$W_{(2,2)}$}
\psfrag{W11h}{$\hat{W}_{(1,1)}$}
\psfrag{W12h}{$\hat{W}_{(1,2)}$}
\psfrag{W21h}{$\hat{W}_{(2,1)}$}
\psfrag{W22h}{$\hat{W}_{(2,2)}$}
\psfrag{WR1}{$W_{(1,1)}, W_{(1,2)},$}
\psfrag{WR2}{$W_{(2,1)}, W_{(2,2)}$}
    \includegraphics[scale=0.45]{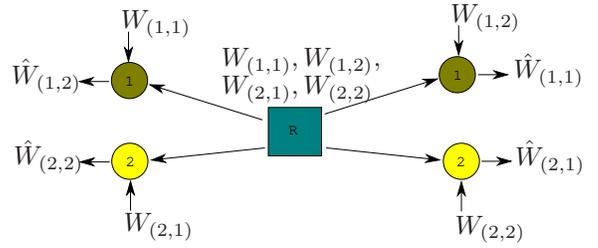}
\end{center}
    \caption{Messages in the two-pair case
    \label{fig:2HopRLayMessage}}
\end{figure}

In this paper, we analyze the broadcasting step by using linear and nonlinear precoding strategies at the relay. For simplicity, we assume that the first hop was successful, i.e. all messages were received with an asymptotically small error probability. This assumption is valid given the rates are within the MAC capacity region in the first hop and the first hop does not pose a bottleneck for the system. The resulting communication scenario is illustrated in Fig.~\ref{fig:2HopRLayMessage} for the two-pair case. In this case, the relay has in total four independent messages, given by $W_{(1,1)}, W_{(1,2)}, W_{(2,1)}$, and $W_{(2,2)}$. Obviously, each node is aware of its own message which was transmitted in the first hop. Furthermore, each node is interested in the message from the other node belonging to the same pair, illustrated by $\hat{W}_{(i,k)}$.

From the information available at the receivers, it becomes clear that this problem distinguishes itself from a multicast setup, in which a common information is broadcasted to a group of receivers without a priori information at the receivers. For the same reason, it is also different from the well studied broadcast channel.
We would also like to emphasize that there are in total $N$ pairs of nodes, i.e. we have $2N$ nodes in total, while the relay is only equipped with $N$ antennas. As a consequence, complete interference avoidance using zero-forcing transmission is not possible.


In the following, we are dealing basically with two optimization problems. In the first optimization problem, we are optimizing a weighted sum-rate subject to a transmit power constraint at the relay. In the second optimization problem, we minimize the total transmitted power from the relay under a minimum $\mathsf{SINR}$ requirement for each node. For both optimization problems we consider a linear as well as a non-linear precoding strategy.

\section{Maximizing sum rate}\label{sec:Rate}
We start with the linear strategy.

\subsection{Linear techniques}

Using linear precoding at the relay, the received signal at node $(i,k)$ is given by
\begin{align}
\tilde{y}_{(i,k)} &=\sqrt{p_i}\mathbf{h}_{(i,k)}^H\mathbf{u}_i s_i+\sum_{j \neq i}\sqrt{p_j}\mathbf{h}_{(i,k)}^H\mathbf{u}_j s_j+{n}_{(i,k)},\\
& \text{for } i=1,\dots,N, \; k=1,2,
\end{align}
where $p_i$ is the power allocation to node-pair $i$, $\mathbf{u}_i$ is the unit-norm beamformer for node-pair $i$, $\mathbf{h}_{(i,k)}$ describes the channel to the $k$-th node, $k={1,2}$, of node-pair $i$.
 The channels between the relay and the nodes are modeled as complex Gaussian random variables, i.e. $\mathcal{CN}(0,1)$.
 The relay has an power constraint $\sum_i p_i\leq P$, $n_{(i,k)}$ is the additive white Gaussian noise with variance $\sigma^2$ at the  $k$-th node of node-pair $i$.
Thus, the transmit signal is of the form
\begin{align}
    \sum_{i}^N\sqrt{p_i}\mathbf{u}_i s_i,
\end{align}
where
 $s_i$ is the complex valued signal denoting the information signal for node-pair $i$.
The received $\mathsf{SINR}_{(i,k)}$ at node $(i,k)$ is
\begin{align}
\mathsf{SINR}_{(i,k)}=\frac{p_i\left|\mathbf{h}_{(i,k)}^H\mathbf{u}_i\right|^2}{\sum_{j \neq i}p_j\left|\mathbf{h}_{(i,k)}^H\mathbf{u}_j\right|^2+\sigma^2}
\end{align}

It has been shown in~\cite{RafaelOechtBiDiBroadISIT,RafaelOechtBiDiBroadSPAWC} for the special case of one pair, that the capacity region can be characterized by its boundary which corresponds to the weighted sum rate given by
\begin{align}\label{eq:CapTobias}
R=\mu_{(1,1)}\log_2C\left(\frac{|\mathbf{h}_{(1,1)}^H\mathbf{u}_1|^2P}{\sigma^2}\right)+\mu_{(1,2)}C\left(\frac{|\mathbf{h}_{(1,2)}^H\mathbf{u}_1|^2P}{\sigma^2}\right),
\end{align}
with weights $\mu_{(1,1)}$ and $\mu_{(1,2)}$, where $C(x)=\log_2(1+x)$.

Thus, using~\eqref{eq:CapTobias} and treating the multi-user interference as noise, the sum rate can be characterized for the general case by
\begin{align}\label{eq:RateRegIntasNoise}
R=\sum_{i=1}^N C\left(\mathsf{SINR}_{(i,1)}\right)+C\left(\mathsf{SINR}_{(i,2)}\right).
\end{align}
For the two pair case the sum rates of each pair are given by
\begin{align}
 R_1 &=C\left(\frac{p_{1}\left|\mathbf{h}_{(1,1)}^H\mathbf{u}_{1}\right|^2}
{p_{2}\left|\mathbf{h}_{(1,1)}^H\mathbf{u}_{2}\right|^2+\sigma^2}\right)+C\left(\frac{p_{1}\left|\mathbf{h}_{(1,2)}^H
\mathbf{u}_{1}\right|^2}{p_{2}\left|\mathbf{h}_{(1,2)}^H\mathbf{u}_{2}\right|^2+\sigma^2}\right) \nonumber\\
R_2 & =C\left(\frac{p_{2}\left|\mathbf{h}_{(2,1)}^H\mathbf{u}_{2}\right|^2}
{p_{1}\left|\mathbf{h}_{(2,1)}^H\mathbf{u}_{1}\right|^2+\sigma^2}\right)+C\left(\frac{p_{2}\left|\mathbf{h}_{(2,2)}^H
\mathbf{u}_{2}\right|^2}{p_{1}\left|\mathbf{h}_{(2,2)}^H\mathbf{u}_{1}\right|^2+\sigma^2}\right) \nonumber.
\end{align}\todo{Without proof: The rate region is probably the union of three rate regions, where the beamformers are similar to the Tobias's approach however with an additional component to the orthogonal directions of the interfering links. Example: For the first region $\mathbf{w}_1$ and similarly for $\mathbf{w}_2$ are a linear combination of the directions of $\mathbf{h}_{1,1}$ and $\mathbf{h}_{1,2}$ and also in the orthogonal subspace of $\mathbf{h}_{2,1}$. For the second region $\mathbf{w}_1$ and similarly for $\mathbf{w}_2$ are a linear combination of the directions of $\mathbf{h}_{1,1}$ and $\mathbf{h}_{1,2}$ and also in the orthogonal subspace of  $\mathbf{h}_{2,2}$. For the third region $\mathbf{w}_1$ and similarly for $\mathbf{w}_2$ are a linear combination of the directions of $\mathbf{h}_{1,1}$ and $\mathbf{h}_{1,2}$ and also a in the orthogonal subspace of the subspace spanned by the channels $\mathbf{h}_{2,1}, \mathbf{h}_{2,2}$.}

In the following paragraph, we are deriving the rate expressions achievable with non-linear precoding, namely dirty paper coding. We then describe the choice of beamforming using non-linear precoding with the goal of maximizing the weighted sum rate subject to power constraints. The differences in the expressions for linear precoding are described along with the discussion on the algorithm.

\subsection{Dirty-Paper Techniques}
With dirty-paper coding~\cite{CostaDirty}, we are able to pre-compensate the interference since it is already known at the transmitter. By assuming an arbitrary but fixed encoding order, the $\mathsf{SINR}_{\pi(i),k}$ of user $k$ of pair $i$ is given by
\begin{align}\label{eq:SINRDirtyPap}
\mathsf{SINR}_{(\pi(i),k)}=\frac{p_{\pi(i)}\left|\mathbf{h}_{(\pi(i),k)}^H\mathbf{u}_{\pi(i)}\right|^2}{\sum_{j > i}p_{\pi(j)}\left|\mathbf{h}_{(\pi(i),k)}^H\mathbf{u}_{\pi(j)}\right|^2+\sigma^2}.
\end{align}


Thus, we have the following rate vector $\mathbf{R}=[R_{(\Sigma,\pi(1))},\dots, R_{(\Sigma,\pi(i))}, \dots, R_{(\Sigma,\pi(N))}]$, where the individual sum rates $R_{(\Sigma,\pi(i))}$ for each pair are given by
\begin{align}\label{eq:RateDirtyPaperTobias}
R_{(\Sigma,\pi(i))}=C\left(\mathsf{SINR}_{(\pi(i),1)}\right)+C\left(\mathsf{SINR}_{(\pi(i),2)}\right).
\end{align}

The dirty rate paper region $R_{\mathsf{DPC}}$ is then given by
\begin{align}
R_{\mathsf{DPC}}=\mathrm{ConvexHull}\left(\bigcup_{\pi, p_i}\mathbf{R}(\pi, p_i)\right),
\end{align}
which is the the convex hull of the union of all rate vectors $\mathbf{R}(\pi, p_i)$ over all powers $p_i$ and over all permutations $(\pi(1),\dots, \pi(N))$.


Assuming that the interference of the other nodes in~\eqref{eq:SINRDirtyPap} and~\eqref{eq:RateDirtyPaperTobias} can be regarded as Gaussian, in our approach the beamforming vectors are computed very efficiently for a given order of precoding in the following successive way.
First of all, from expression~\eqref{eq:SINRDirtyPap}, we observe that the node pair $\pi(N)$ encoded last experiences no interference from the other nodes.

Thus, using the approach in~\cite{RafaelOechtBiDiBroadSPAWC} the beamforming vector for the node pair $\pi(N)$ is obtained by
\begin{align}
\mathbf{u}_{\pi(N)}(t)=\frac{t\mathbf{g}_{(\pi(N),1)}+(1-t)\exp(-\jmath \phi)\mathbf{g}_{(\pi(N),2)}}{||t\mathbf{g}_{(\pi(N),1)}+(1-t)\exp(-\jmath \phi)\mathbf{g}_{(\pi(N),2)}||}\label{eq:BeamformSinglePair}
\end{align}
with $\mathbf{g}_{(\pi(N),1)}=\mathbf{h}_{(\pi(N),1)}/||\mathbf{h}_{(\pi(N),1)}||$, $\mathbf{g}_{(\pi(N),2)}=\mathbf{h}_{(\pi(N),2)}/||\mathbf{h}_{(\pi(N),2)}||$, and $\phi=\arg \left(\mathbf{g}_{(\pi(N),1)}^H\mathbf{g}_{(\pi(N),2)}\right)$. The weight $t$ can be used to prioritize one of the nodes.
Intuitively, the beamforming vector in~\eqref{eq:BeamformSinglePair} represents a linear combination (up to a phase adjustment) of maximum-ratio transmission beamformers in the direction of the node pairs.

The node pair $\pi(N-1)$ is considered next. The node pair $\pi(N-1)$ observes interference from the beam intended for the node pair $\pi(N)$, which is exactly known.
Assuming again that the interference from the node-pair $\pi(N)$ can be regarded as additional white Gaussian noise then the overall interference plus noise (referred to as effective noise in the following) is distributed as $\mathcal{CN}(0,\sigma_{(\pi(N-1),k)}^2)$ with
\begin{align}
\sigma_{(\pi(N-1),k)}^2=p_{\pi(N)}\left|\mathbf{h}_{(\pi(N-1),k)}^H\mathbf{u}_{\pi(N)}\right|^2+\sigma^2\label{eq:EffNoiseDirty},
\end{align}
we compute the beamforming vector for node pair $\pi(K-1)$ very efficiently as in~\eqref{eq:BeamformSinglePair}.

This process is continued until $p_{\pi(1)}$ is determined.
The above algorithm has to be repeated for all possible user orderings $\pi$. In the linear precoding case, each receiver is observing interference from all other beams in the system, thus the effective noise variance~\eqref{eq:EffNoiseDirty} is changed accordingly. The beamformers are obtained by evaluating~\eqref{eq:BeamformSinglePair} accordingly.
Note that there is no claim of optimality in terms of achievable rates of the beamforming and precoding approaches discussed above.
Due to the non-convex structure of the optimization problem, an optimal beamforming strategy for the weighted rate maximization is difficult to obtain. However, note that beamforming has always the advantage of simple processing at the transmitter (scalar instead of vector coding) and receiver (single stream decoding) and is thus often considered in wireless standards such as LTE and WiMAX, which justifies the analysis of such schemes.
As an alternative to the approach discussed above, in the following section the equivalent problem of power minimization subject to quality-service-constraints (QoS) in terms of required $\mathsf{SINR}$ is considered, first for the linear precoding case, followed by non-linear precoding.

\section{Minimizing power}\label{sec:Power}

\subsection{Linear techniques}
\begin{figure}
  \subfigure[Downlink]{\includegraphics[scale=0.75]{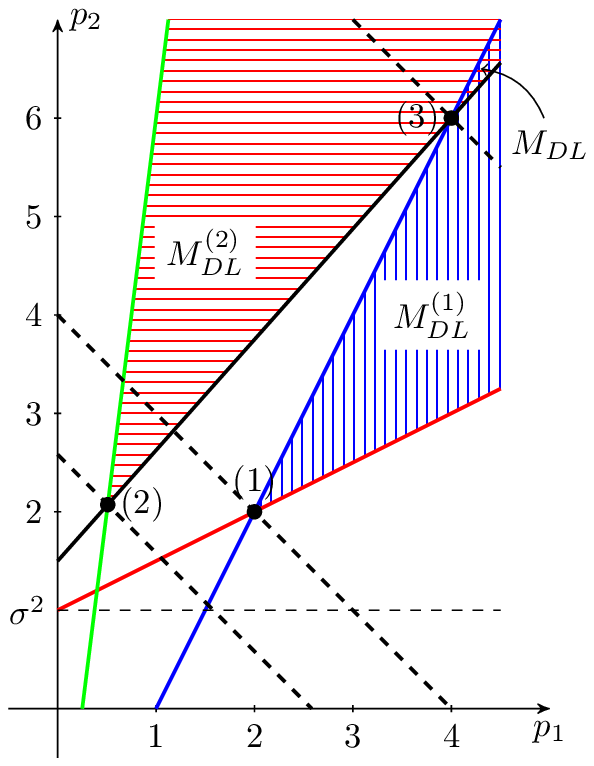}}
  \subfigure[Dual Uplink]{\includegraphics[scale=0.75]{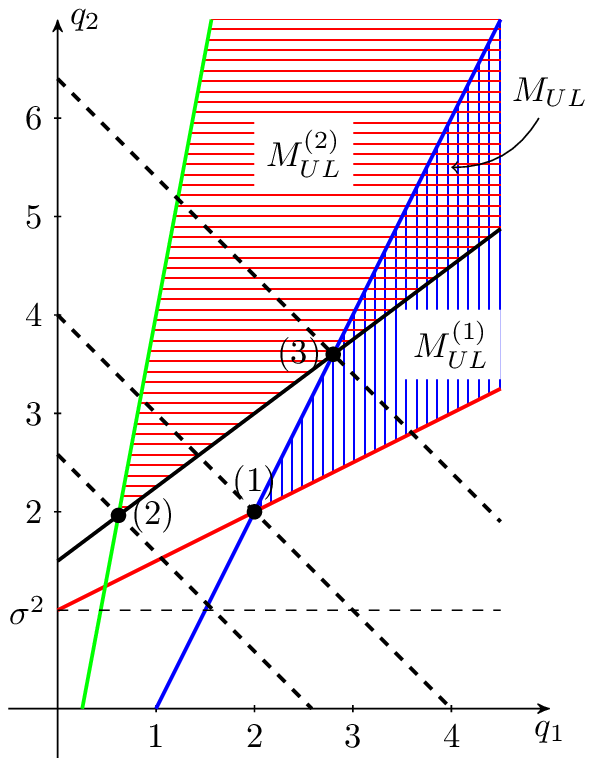}}
  \caption{Illustration of counter-example to uplink-downlink duality}\label{fig:UplinkDownlinDual}
\end{figure}

In the following, we will show that the uplink-downlink duality~\cite{SchubertBoche,ViswanathTse} does not hold in the setup considered here.
We start with the downlink.
The optimization problem by using linear precoding is given by
\begin{align}
\text{min.} & \sum_{i=1}^{N} p_i \label{eq:MinPower}
\\
    \text{s.t.} &\; \exists\, \mathbf{u}_{1},\dots, \mathbf{u}_{N}, ||\mathbf{u}_i||^2=1, 1\leq l \leq N, \nonumber \\
    &\; \frac{p_i\left|\mathbf{h}_{(i,k)}^H\mathbf{u}_{i}\right|^2}{\sum\limits_{j=1, j\neq i} \left|\mathbf{h}_{(i,k)}^H\mathbf{u}_{j}\right|^2p_j+\sigma^2}\geq \gamma_{(i,k)}, \nonumber\\
    & \hspace{4cm} k=1,2, i=1,\dots, N \nonumber
\end{align}
where $\gamma_{(i,k)}$ is the $\mathsf{SINR}$ requirements at node $(i,k)$.
Let us introduce the following matrices $\mathbf{V}^{(k)}$ with
\begin{align}
&    \mathbf{V}_{i,j}^{(k)}=|\mathbf{h}_{(i,k)}^H\mathbf{u}_j|^2, \quad 1\leq i,j \leq N, k=1,2 \\
&  \mathbf{V}_{i,i}^{(k)}=0, \quad 1 \leq i \leq N.
\end{align}
Note that the $\mathbf{V}^{(k)}$ are functions of the beamforming vectors, i.e. $\mathbf{V}^{(k)}=\mathbf{V}^{(k)}\left(\mathbf{U}\right)$, with $\mathbf{U}=\left[\mathbf{u}_1, \dots, \mathbf{u}_N \right]$.
Then~\eqref{eq:MinPower} can be rewritten in the following form
\begin{align}
\text{min.} & \sum_{i=1}^{N} p_i \label{eq:MinPowerRew}
\\
    \text{s.t.} &\; \exists\, \mathbf{U}, ||\mathbf{u}_i||^2=1, 1\leq l \leq N \nonumber \\
    &\; \mathbf{p}\geq \left(\mathbf{D}^{(k)}\right)^{-1}\left(\mathbf{\Gamma}^{(k)}\mathbf{V}^{(k)}\mathbf{p}+\sigma^2\mathbf{\Gamma}^{(k)}\mathbf{\mathds{1}}\right)\; k=1,2 \nonumber \\
    & \mathbf{D}^{(k)}=\mathrm{diag}\left(|\mathbf{h}_{(i,k)}^H\mathbf{u}_{i}|^2\right) \nonumber,
\end{align}
with $\mathbf{\Gamma}^{(k)}=\mathrm{diag}\left(\gamma_{(1,k)},\dots, \gamma_{(N,k)}\right)$ representing the matrix of $\mathsf{SINR}$ requirements, where $\mathbf{\mathds{1}}$ is the all one vector.
The power minimization problem in the \emph{dual} uplink is given by
\begin{align}
\text{min.} & \sum_{i=1}^{N} q_i \label{eq:MinPowerRewUp}
\\
    \text{s.t.} &\; \mathbf{q}\geq \left(\mathbf{D}^{(k)}\right)^{-1}\left(\mathbf{\Gamma}^{(k)}\left(\mathbf{V}^{(k)}\right)^H\mathbf{q}+\sigma^2\mathbf{\Gamma}^{(k)}\mathbf{\mathds{1}}\right)\; k=1,2 \nonumber
\end{align}
Note that at the optimum the constraints are achieved with equality (otherwise, the power corresponding to the constraint for $\gamma_{(i,k)}$ can be reduced such that equality is obtained, which would reduce the power spent and thus would be a contradiction.)
In the following, by using a counter-example we show that uplink-downlink-duality does not hold. Consider the case $N=2$, with $\
\mathbf{D}^{(k)}=\mathbf{I}$, $k=1,2$, $\mathbf{\Gamma}^{(1)}=\mathbf{I}$, and $\mathbf{\Gamma}^{(2)}\neq \mathbf{I}$.
Then, define the following sets according to the constraints in~\eqref{eq:MinPowerRew} and~\eqref{eq:MinPowerRewUp} for this case
{\small{\begin{align}
    & \mathcal{M}_{\mathrm{DL}}=\left\{\mathbf{p} \in \mathds{R}_+^2: \mathbf{p}\geq \mathbf{\Gamma}^{(k)}\mathbf{V}^{(k)}\mathbf{p} +\sigma^2\mathbf{\Gamma}^{(k)}\left[
                                                                                                                         \begin{smallmatrix}
                                                                                                                           1 \\
                                                                                                                           1 \\
                                                                                                                         \end{smallmatrix}
                                                                                                                       \right], k=1,2
    \right\} \\
   &  \mathcal{M}_{\mathrm{UL}}=\left\{\mathbf{q} \in \mathds{R}_+^2: \mathbf{q}\geq \mathbf{\Gamma}^{(k)}(\mathbf{V}^{(k)})^{H}\mathbf{q} +\sigma^2\mathbf{\Gamma}^{(k)}\left[
                                                                                                                         \begin{smallmatrix}
                                                                                                                           1 \\
                                                                                                                           1 \\
                                                                                                                         \end{smallmatrix}
                                                                                                                       \right], k=1,2
    \right\}\nonumber
\end{align}}}
for downlink and uplink, respectively. Note that $\mathcal{M}_{\mathrm{DL}}=\cap_{k}\mathcal{M}_{\mathrm{DL}}^{(k)}$ and similarly for the uplink. The sets are illustrated in Fig.~\ref{fig:UplinkDownlinDual}.
In Fig.~\ref{fig:UplinkDownlinDual}(a) (analogously in Fig.~\ref{fig:UplinkDownlinDual}(b)) there are three relevant intersections, marked with a number $(1),(2)$, and $(3)$ respectively. The numbers $(1)$ and $(2)$ indicate the minimum of  the subset $M_{\mathrm{DL}}^{(1)}$ and $M_{\mathrm{DL}}^{(2)}$, respectively. The point $(3)$ indicates the minimum of the $M_{\mathrm{DL}}$. For simplicity, it is assumed that $\mathbf{V}^{(1)}$ is symmetric (cf. Fig.~\ref{fig:UplinkDownlinDual}). It follows that $\mathcal{M}_{\mathrm{DL}}^{(1)}$ is symmetric and so is $\mathcal{M}_{\mathrm{UL}}^{(1)}$.  Note that the minimum of $\mathcal{M}_{\mathrm{DL}}^{(1)}$ and $\mathcal{M}_{\mathrm{UL}}^{(1)}$ are on the
same dotted slope~$-1$ line passing through $(1)$ due to duality (since $\sum_i p_i = \sum_i q_i$). The same holds for $\mathcal{M}_{\mathrm{DL}}^{(2)}$ and $\mathcal{M}_{\mathrm{UL}}^{(2)}$. However, as can be seen in the figure it holds not true for $\mathcal{M}_{\mathrm{DL}}$ and $\mathcal{M}_{\mathrm{UL}}$ (compare the dotted lines passing through the points $(3)$), unless
$\mathbf{\Gamma}^{(2)}= \mathbf{I}$ as well, which is in general not the case.

Thus, uplink-downlink duality can not be exploited in order to get a solution for the downlink beamformers as in~\cite{SchubertBoche,ViswanathTse} for the broadcast channel (unless the $\mathsf{SINR}$ requirements $\mathbf{\Gamma}^{(k)}$ are equal for all nodes, which is a case of limited interest).

However, the optimization problem in~\eqref{eq:MinPower} can be solved approximately by using a semidefinite relaxation.
Let us define $\mathbf{Q}_i=p_i\mathbf{u}_i\mathbf{u}_i^H$. In the original problem, the $\mathbf{Q}_i$, $\forall i$, are rank-constraint. This non-convex constraint can be replaced by a convex constraint $\mathbf{Q}_i\succeq \mathbf{0}$, where the notation $\mathbf{A}\succeq \mathbf{B}$ means $\mathbf{A}-\mathbf{B}$ is a positive semidefinite matrix. While the original constraint was of rank $1$, the $\mathbf{Q}_i$ can be of any rank with this relaxation. Thus, the relaxed problem is a semidefinite programming problem (SDP) given by
\begin{align}
\text{minimize} & \sum_{i=1}^{N} \mathrm{trace}\left(\mathbf{Q}_i\right) \label{eq:RelaxedOptMimSNRZF}
\\
    \text{subject to} &\; \frac{\mathbf{h}_{(i,k)}^H\mathbf{Q}_i\mathbf{h}_{(i,k)}}{\sum_{j=1, j\neq i}\mathbf{h}_{(i,k)}^H\mathbf{Q}_{j}\mathbf{h}_{(i,k)}+\sigma^2}\geq \gamma_{(i,k)}\\
    & \; \mathbf{Q}_i\succeq \mathbf{0}.
\end{align}


 The solution to the relaxed SDP~\eqref{eq:RelaxedOptMimSNRZF} gives a lower bound of the objective function and a relaxed solution $\mathbf{Q}_i^*$. However, if that solution $\mathbf{Q}_i^*$ is of rank~$1$, then the solution to the relaxed optimization problem is identical to the solution of the original problem. In general, however, the solution will not have rank~$1$. As an heuristic but efficient approach, here the eigenvectors corresponding to the largest eigenvalue of $\mathbf{Q}_i^*$ are used as a rank~$1$ approximation.

\subsection{Dirty-paper techniques}
The procedure here is similar to the linear beamforming case, except that the order of encoding is an additional parameter which has to be taken into account in the optimization. The SDP relaxation in the case of non-linear precoding is given by
\begin{align}
\text{minimize} & \sum_{i=1}^{N} \mathrm{trace}\left(\mathbf{Q}_i\right) \label{eq:RelaxedOptMimSNRDPC}
\\
    \text{subject to} &\; \frac{\mathbf{h}_{(\pi(i),k)}^H\mathbf{Q}_{\pi(i)}\mathbf{h}_{(\pi(i),k)}}{\sum_{j> i}\mathbf{h}_{(\pi(i),k)}^H\mathbf{Q}_{\pi(j)}\mathbf{h}_{(\pi(i),k)}+\sigma^2}\geq \gamma_{(\pi(i),k)}\\
    & \; \mathbf{Q}_{\pi(i)}\succeq \mathbf{0}.
\end{align}
Once the optimal beamformers are obtained from~\eqref{eq:RelaxedOptMimSNRDPC}, a rank~$1$ approximation is obtained using the same procedure as in the linear case. By defining the $\mathsf{SINR}$ requirements $\gamma_{(\pi(i),k)}$ in terms of rates as follows
\begin{align}
    \gamma_{(\pi(i),k)}=2^{\mu_{\pi(i)}R_{\Sigma, \pi(i)}}-1
\end{align}
with $\sum_i \mu_{i}=1$, we can apply an iterative (bisection)~\cite{BoydVanderberghe} power minimization problem to get an achievable rate region. The parameter  $R_{\Sigma, \pi(i)}$ is changed at each iteration, i.e. the value of $R_{\Sigma, \pi(i)}$ is incremented in the next iteration if~\eqref{eq:RelaxedOptMimSNRDPC} is feasible otherwise decreased.  The step size is obtained by bisection between the last feasible value of $R_{\Sigma, \pi(i)}$, which is initially equal to zero, and the last infeasible value of $R_{\Sigma, \pi(i)}$, which can be chosen as twice the capacity in the single-pair case given in~\eqref{eq:CapTobias} for initialization. The iterations continue until a predefined accuracy is achieved.

\section{Illustration}\label{sec:Illus}
\begin{figure}
\begin{center}
  \psfrag{Sum Rate Pair 2}{\hspace*{-1cm}$\text{Sum-Rate Pair 2}\, R_{\Sigma, 2}$}
    \psfrag{Sum Rate Pair 1}{$\text{Sum-Rate Pair 1}\, R_{\Sigma, 1}$}
  \includegraphics[scale=0.35]{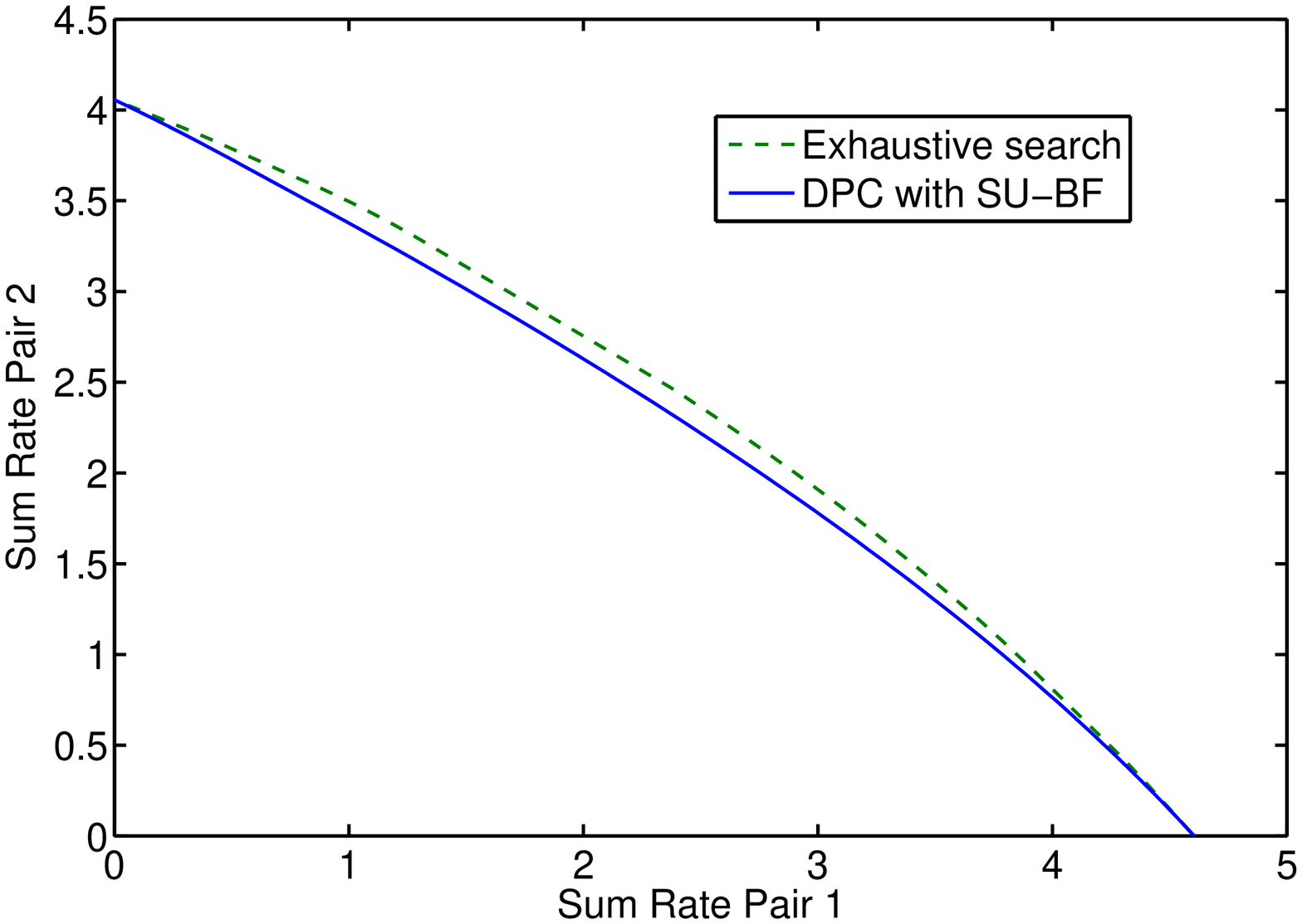}\\
  \caption{Rate Regions achieved with Dirty-Paper Coding; single-user beamforming approach (cf.~\eqref{eq:BeamformSinglePair}) vs. randomly generated beams at $\mathsf{SNR}=3$~dB}\label{fig:RateRegions3dB}
  \end{center}
\end{figure}

In this section, the results are illustrated by means of numerical simulations. Due to the similarity of the results, we focus in this section on the non-linear precoding strategy, i.e. dirty-paper coding.
In Fig.~\ref{fig:RateRegions3dB}, the rate region of a two pair network is shown for a $\mathsf{SNR}=3$~dB. Note that on the coordinate we have the sum rate of pair~$i=1$, while on the ordinate we have the sum rate of pair~$i=2$. Thus, the figure is a projection of a four-dimensional rate region to a two-dimensional one. For both curves in the plot, dirty-paper coding is used. As beamforming strategy the single-pair beamforming approach (cf.~\eqref{eq:BeamformSinglePair}) is used and compared to an exhaustive search over a huge set of randomly generated beams.
The channel vectors used in the plots have the following norms: $||\mathbf{h}_{11}||^2_2=3.28, ||\mathbf{h}_{12}||^2_2=2.9, ||\mathbf{h}_{21}||^2_2=1.77$, and $||\mathbf{h}_{22}||^2_2=2.2$. From the figure, we observe that the single-pair approach performs very close to the exhaustive search.

\begin{figure}
  \begin{center}
  \psfrag{Sum  Rate Pair 2}{\hspace*{-1cm}$\text{Sum-Rate Pair 2}\, R_{\Sigma, 2}$}
    \psfrag{Sum Rate Pair 1}{$\text{Sum-Rate Pair 1}\, R_{\Sigma, 1}$}
  \includegraphics[scale=0.35]{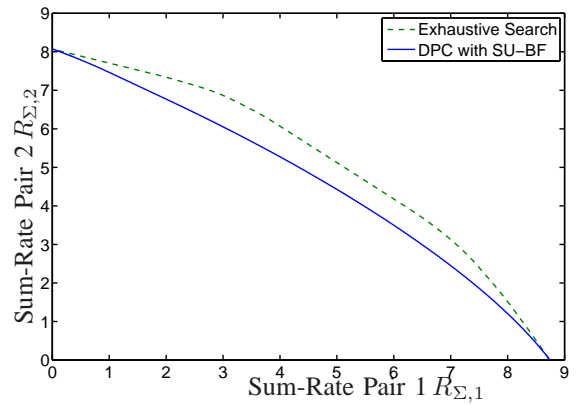}\\
  \caption{Rate Regions achieved with Dirty-Paper Coding; single-user beamforming approach (cf.~\eqref{eq:BeamformSinglePair}) vs. randomly generated beams at $\mathsf{SNR}=10$~dB}\label{fig:RateRegions10dB}
  \end{center}
\end{figure}

The same strategies are compared in Fig.~\ref{fig:RateRegions10dB}, now for $10$~dB. Here, we observe a small gap between the single-pair approach and the exhaustive search. Still, the single-user strategy performs reasonably well.

The next Fig.~\ref{fig:RateRegions30dB} shows the performance of the two strategies for $30$~dB. In addition to this, the region obtained using the relaxation method given in~\eqref{eq:RelaxedOptMimSNRDPC} is plotted.
From the figures, we observe that the gap between the exhaustive search and the single-pair approach has increased further. The region obtained using the relaxation method is comparable to the one using the single-pair approach.
From the figures, we conclude that our approach is close to the exhaustive search for low and average $\mathsf{SNR}$, but is suboptimal for high $\mathsf{SNR}$.

\begin{figure}
\begin{center}
  \psfrag{Sum Rate pair 2}{\hspace*{-1cm}$\text{Sum-Rate Pair 2}\, R_{\Sigma, 2}$}
    \psfrag{Sum Rate Pair 1}{$\text{Sum-Rate Pair 1}\, R_{\Sigma, 1}$}
  \includegraphics[scale=0.35]{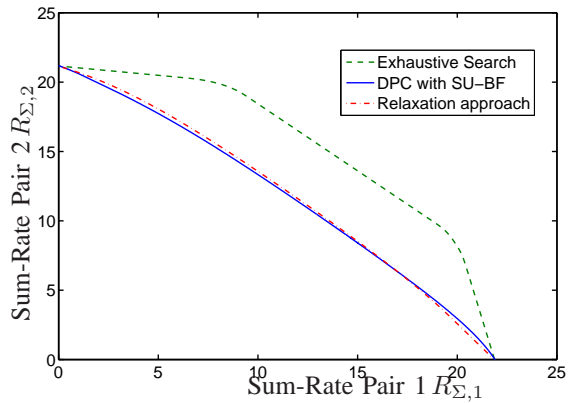}\\
  \caption{Rate Regions achieved with Dirty-Paper Coding; single-user beamforming approach (cf.~\eqref{eq:BeamformSinglePair}), randomly generated beams, and the region obtained (dashed-dotted line) using the relaxation method (cf.~\eqref{eq:RelaxedOptMimSNRDPC}) at $\mathsf{SNR}=30$~dB}\label{fig:RateRegions30dB}
  \end{center}
\end{figure}

\section{Conclusion}\label{sec:Conc}
In this paper, we have addressed the problem of a multi-user relay network, where multiple single-antenna node pairs want to exchange information by using a multiple antenna relay node. In the first step, the nodes transmit their data to the relay, while in the second step, the relay is broadcasting the data by using linear and non-linear precoding strategies. We focused on the broadcasting step and first considered the problem of maximizing the overall rate achievable using linear and dirty-paper type precoding strategies at the relay. Then, we considered the problem of minimizing the total power at the relay subject to individual $\mathsf{SINR}$ constraints. We showed that the downlink-uplink duality does not hold for the setup considered here. We also showed that using the beamforming strategy which is optimal in the single-pair case performs very well for practically relevant values of $\mathsf{SNR}$.

{
\bibliographystyle{IEEE}

}

\end{document}